\documentclass[epj,nopacs,fleqn]{svjour}

\usepackage{graphicx}
\usepackage{amsmath}
\usepackage{amssymb}
\usepackage{color}

\usepackage{listings}
\newcommand{\helas}{{\tt HELAS}\ }
\newcommand{\gmn}{g_{\mu\nu}}
\newcommand{\gmnu}{g^{\mu\nu}}

\newcommand{\gmntu}{\tilde g^{\mu\nu}}
\newcommand{\lam}{\lambda}
\newcommand{\sig}{\sigma}
\newcommand{\alp}{\alpha}
\newcommand{\bet}{\beta}
\newcommand{\eps}{\epsilon}
\newcommand{\ept}{\tilde\epsilon}
\newcommand{\GeV}{\mbox{GeV}}
\newcommand{\TeV}{\mbox{TeV}}
\newcommand{\pT}{\mbox{$p_T$}}
\newcommand{\etapTmin}{\mbox{$\eta(p_{T\mathrm{min}})$}}

\newcommand{\kTij}{\mbox{$k_{T ij}$}}

\newcommand{\kTijmin}{\mbox{$k_{T ij\,\mathrm{min}}$}}

\begin{document}

\title{QED and QCD helicity amplitudes in parton-shower gauge}

\author{Kaoru Hagiwara
	\inst{1,2}\thanks{kaoru.hagiwara@kek.jp}
	\and	 
	Junichi Kanzaki\inst{3,4}\thanks{junichi.kanzaki@ipmu.jp}
	\and
	Kentarou Mawatari\inst{5}
\thanks{mawatari@iwate-u.ac.jp}%
}                     
%
%
\institute{
	Institute of Science and Engineering, Shimane University, Matsue 690-8504, Japan
	\and
	KEK Theory Center, Tsukuba 305-0801, Japan
	\and 
 	KEK and Sokendai, Tsukuba 305-0801, Japan
  \and
  Address since April 2020: Kavli Institute for the Physics and Mathematics of the Universe, The University of Tokyo, \\ Kashiwa 277-8583, Japan
	\and
	Faculty of Education, Iwate University, Morioka, Iwate 020-8550, Japan
	 }

\date{}

\abstract{
We introduce photon and gluon propagators in which the scalar polarization component is subtracted systematically by making use of the BRST invariance of the off-shell vector boson created from physical on-shell states.
The propagator has the light-cone gauge form, where the spacial component of the gauge vector points along the negative of the off-shell vector boson momentum.
We call the gauge as parton-shower gauge, since
in collinear configurations
the absolute value squared of each Feynman amplitude reproduces all the singular behaviors of the corresponding parton shower in this gauge.
We introduce new {\tt HELAS} codes that can be used to calculate the tree-level helicity amplitudes of arbitrary QED and QCD processes by using {\tt MadGraph}.
The absence of subtle gauge cancellation among Feynman amplitudes allows numerical codes to evaluate singular behaviors accurately, and helps us gaining physical insights on interference patterns.
%
}

\maketitle

\vspace*{-11cm}
\noindent KEK-TH-2182
\vspace*{9.15cm}

\section{Introduction}\label{intro}

\helas ({\tt HEL}icity {\tt A}mplitude {\tt S}ubroutines) is a set of Fortran subroutines which enable us to compute the helicity amplitudes of an arbitrary tree-level Feynman diagram~\cite{Hagiwara:1990dw,Murayama:1992gi}.
The code has been used in various event generators such as {\tt MadGraph5\_aMC@NLO} ({\tt MG5aMC})~\cite{Alwall:2014hca}.%
\footnote{
While {\tt MadGraph/MadEvent v4}~\cite{Alwall:2007st} employs the original \helas library, {\tt MG5aMC} adopts {\tt ALOHA}~\cite{deAquino:2011ub}, which automatically generates the \helas library by using the {\tt UFO} format~\cite{Degrande:2011ua}.}

\helas employs the Feynman-gauge propagator for massless gauge bosons.
It often leads to subtle gauge cancellation among Feynman amplitudes, that sometimes give rise to a total loss of accuracy even with double precision.
For example, the size of cancellation among amplitudes for $e^{+} e^{-} \rightarrow e^{+} e^{-} Z$ at 2~\TeV\ collision  amounts up to $\sim10^{12}$ and that results in about 10\% inaccuracy of the calculation of the total cross section~\cite{Hagiwara:1990gk}.

In this article, 
we propose a different gauge choice for the propagator for massless gauge bosons,
in which the scalar polarization component is subtracted systematically by making use of the BRST invariance of the off-shell vector boson created from physical on-shell states.
The propagator has the light-cone gauge form, and
we may call the gauge as parton-shower (PS) gauge, since, as shown below, the absolute value squared of each Feynman amplitude reproduces all the singular behaviors of the corresponding parton shower in this gauge.%
\footnote{After completion of the paper, we found that the same form of the gluon propagator is used indeed in a parton-shower program by Nagy and Soper~\cite{Nagy:2007ty,Nagy:2014mqa}.}

After introducing the PS gauge in Sect.~\ref{sec:psg},
we discuss the relation between the off-shell vector boson currents and the parton splitting amplitudes in Sect.~\ref{sec:splitting}.
In Sect.~\ref{sec:results} we give sample numerical results 
to demonstrate that calculations in the PS gauge can avoid subtle gauge cancellations among Feynman amplitudes by comparing with those in the Feynman gauge.
Sect.~\ref{sec:summary} presents our brief summary.
In \ref{sec:codes} we give new \helas subroutines in the PS gauge.

\section{Parton-shower gauge}\label{sec:psg}

In this section, we introduce PS gauge for massless gauge-boson propagators.

The numerator for a massless gauge boson propagator in a covariant gauge can be decomposed as
\begin{align}
  -\gmnu+(1-\xi)\frac{q^\mu q^\nu}{q^2}
 &=\sum_{\lam=\pm,0}\eta_\lam\,\eps^\mu(q,\lam)\,\eps^\nu(q,\lam)^* \notag\\
  &\quad-\xi\,\eta_{S}\,\eps^\mu(q,S)\,\eps^\nu(q,S)^*,
  \label{prop}
\end{align}
where $\xi$ is a gauge parameter;
$\eta_\lam\!=\!\eta_{S}\!=\!1$ when $q^2>0$, while $\eta_\lam=(-1)^{\lam+1}$ and $\eta_{S}\!=\!-1$ when $q^2<0$;
$\eps^{\mu}(q,\lam)$ is the polarization vector with momentum $q$ and helicity $\lam$,
while $\eps^{\mu}(q,S)$ is the scalar polarization state
\begin{align}
  \eps^\mu(q,S)=\frac{q^{\mu}}{\sqrt{|q^2|}}.   
\label{epsS}
\end{align}
The polarization vectors for the three helicity states satisfy
\begin{align}
  q_\mu\eps^\mu(q,\lam)=0,
\end{align}
and are normalized as
\begin{align}
  \eps_\mu(q,\lam)\,\eps^\mu(q,\lam')^*=-\delta_{\lam\lam'}.
\end{align}
The gauge invariance of the scattering amplitudes in QED and QCD tells that the term proportional to $q^{\mu}q^{\nu}$ do not affect the amplitudes, and we can express the propagator in terms of the three polarization vectors as
\begin{align}
  -\gmnu+\frac{q^{\mu}q^{\nu}}{q^2}
 =\sum_{\lam=\pm,0}\eta_\lam\,\eps^{\mu}(q,\lam)\,\eps^{\nu}(q,\lam)^*. 
\label{comp}
\end{align}
We note here that the longitudinally polarized vector-boson wave function can be decomposed into two parts
\begin{align}
  \eps^{\mu}(q,0) = \ept^{\mu}(q,0) + \eps^{\mu}(q,S),
\end{align}
where at high energies ($|q^{2}|/|{\vec q}|^{2} \ll 1$) the $\ept^{\mu}(q,0)$ term is suppressed.
The longitudinal polarization contribution to the propagator \eqref{comp} hence has four components:
\begin{align}
  \eps^{\mu}(q,0)\eps^{\nu}(q,0)^* 
 &=  \ept^{\mu}(q,0)\ept^{\nu}(q,0)^* 
    +\ept^{\mu}(q,0)\eps^{\nu}(q,S)^* \notag \\
 & +\eps^{\mu}(q,S)\ept^{\nu}(q,0)^* 
    +\eps^{\mu}(q,S)\eps^{\nu}(q,S)^*.
 \label{epseps}
\end{align}
The last term does not contribute because it has the same form as the gauge term in~\eqref{prop}.
In fact, neither the second nor the third term contributes to the scattering amplitudes, because the BRST invariance of the quantum gauge field theory~\cite{Becchi:1975nq,Tyutin:1975qk} ensures that the gauge fixing condition annihilates the physical states:
\begin{align}
  \langle\mathrm{phys}|\partial_{\mu} A^{\mu} |\mathrm{phys}\rangle = 0,  
\label{brst}
\end{align}
for the covariant gauge fixing.
Because the scattering amplitudes are transitions among physical states, the scalar polarization component $\eps^{\mu}(q,S)$ in all the propagators of the tree-level amplitudes annihilates the physical states that gives rise to the off-shell vector boson current.
In short, only the first term in the right-hand side of eq.~\eqref{epseps} contributes to the scattering amplitudes at the tree-level in QED and QCD.

We now introduce a tensor
\begin{align}
  \gmntu=\gmnu-\frac{n^\mu q^\nu+q^\mu n^\nu}{n\cdot q}
\end{align}
with
\begin{align}
  n^\mu=({\rm sgn}(q^0),-\vec q/|\vec q|),
\label{nvec}
\end{align}
and obtain the projected polarization vector as
\begin{align}
  \ept^{\mu}(q,\lam)=\gmntu\eps_{\nu}(q,\lam).
\end{align}
Here the transverse polarization states remain the same, 
while the longitudinal state is subtracted by the scalar state:
\begin{align}
  \ept^{\mu}(q,\pm)&=\eps^{\mu}(q,\pm), \\
  \ept^{\mu}(q,0)  &=\eps^{\mu}(q,0)-\eps^{\mu}(q,S),
\end{align}
for all $q^\mu$.
The BRST invariance~\eqref{brst} tells that only the projected polarization vectors contribute,
and the propagator can be expressed as
\begin{align}
  \sum_{\lam=\pm,0}\eta_\lam\,\ept^\mu(q,\lam)\,\ept^\nu(q,\lam)^* 
 &=\tilde g^{\mu\alpha}\tilde g^{\nu\beta}
   \Big(-g_{\alpha\beta}+\frac{q_\alpha q_\beta}{q^2}\Big) \notag\\
 &=-\gmnu+\frac{n^\mu q^\nu+q^\mu n^\nu}{n\cdot q} \notag\\
 &=-\gmntu.
\label{psg}
\end{align}
It has the form of the light-cone gauge propagator,%
\footnote{The light-cone gauge propagator with the gauge vector~\eqref{psg} was introduced in ref.~\cite{Chen:2016wkt} in order to obtain splitting amplitudes for massive vector bosons in the standard model.
In spontaneously broken gauge theories, the BRST invariance~\eqref{brst} takes the form $\langle\mathrm{phys}| (\partial_{\mu} V^{\mu}+\xi m_V\chi_V) |\mathrm{phys}\rangle = 0$, relating the scalar polarization of the massive vector boson and the corresponding Goldstone boson contributions. The gauge is named Goldstone-equivalence gauge because the dominant piece of the longitudinal polarization contribution is replaced by the Goldstone boson exchange amplitudes. One can obtain numerical codes that replaces the unitary gauge propagators by the sum of the light-cone gauge propagator of the vector boson and the Goldstone boson~\cite{Chen:2020}.
}
but we should note that the gauge vector \eqref{nvec} depends on the three momentum of the virtual vector boson.
We call the propagator form \eqref{psg} as the ``parton-shower (PS)" gauge because of the property of the splitting amplitudes calculated in this method,
as shown in the next section.

\section{Relation to the splitting amplitudes}\label{sec:splitting}

In this section, we discuss the relation between the off-shell vector boson currents and the parton splitting amplitudes. 
 
We consider an amplitude for processes with initial state radiations (space-like; $q^2<0$) that produce a nearly on-shell virtual gluon
\begin{align}
  q(k,\sig) &\to q(k',\sig')  + g(q)^*, \\
  g(k,\sig) &\to g(k',\sig') + g(q)^*,
\end{align}
with the four momentum $q=k-k'$.
$\sig$ and $\sig'$ denote the initial and final parton helicities, respectively.
The amplitudes for the full process can be expressed as
\begin{align}
  {\cal M}=J^\mu(k,k';\sig,\sig')\frac{-\gmn}{q^2}T^\nu
\end{align}
with
\begin{align}
  J^\mu(k,k';\sig,\sig') = g_st^a\,\bar{u}(k',\sig') \gamma^\mu u(k,\sig)
\label{J_q}  
\end{align}
for $q\to qg$ splitting, and
\begin{align}
  J^\mu(k,k';\sig,\sig')
 =g_sf^{abc}\,\Gamma^{\mu\alp\bet}\, \eps_\alp^a(k,\sig)\, \eps_\bet^b(k',\sig')^*
\label{J_g}   
\end{align}
for $g\to gg$ splitting.  
The splitting currents are then expanded by using the polarization vectors with three helicities along the momentum transfer ($q$) direction:
\begin{align}
  &J^\mu(k,k';\sig,\sig') \notag\\
  &=\Big(g^{\mu\nu} -\frac{q^\mu q^{\nu}}{q^2}\Big)\, J_\nu(k,k';\sig,\sig') \notag\\
  &=\sum_{\lam=\pm,0}(-\eta_\lam)\,\eps^{\mu}(q,\lam)\,\eps^{\nu}(q,\lam)^*
  J_\nu(k,k';\sig,\sig') \notag\\
  &=\sum_{\lam=\pm,0} {\cal J}_{\sig\sig'}^\lam(k,k') \,\eps^\mu(q,\lam),  
\label{J_fg}  
\end{align}
where the splitting amplitudes~\cite{Hagiwara:2009wt} are introduced as
\begin{align}
   {\cal J}_{\sig\sig'}^\lam(k,k') = -\eta_\lam \,J_\mu(k,k';\sig,\sig')\,\eps^\mu(q,\lam)^{*}. 
\label{amp}  
\end{align}
The $q\to qg$ amplitudes are invariant under the boost along the virtual gluon momentum ($q$) direction,
while the $g\to gg$ ones are not always invariant because of possible gauge choice of the polarization vectors for the external massless gauge bosons.
By adopting the common light-cone gauge \eqref{nvec} for all the external and the off-shell gluons, the $g\to gg$ amplitudes can also be boost invariant along the current momentum direction~\cite{Hagiwara:2009wt}.
By using the boost-invariant splitting amplitudes ${\cal J}_{\sig\sig'}^\lam(k,k')$ of ref.~\cite{Hagiwara:2009wt} and by making the on-shell approximation for the virtual gluon in the hard scattering amplitudes,
the azimuthal angle correlations between the initial state splitting ($q\to qg^*$ and $g\to gg^*$) plane and the hard scattering plane are reproduced in the collinear limit.
For off-shell gluons, however, the longitudinally polarized vector $\eps^\mu(q,0)$ in the right-hand side of eq.~\eqref{J_fg} behaves badly in the boosted frame, which often leads to severe numerical cancellation among amplitudes in the exact matrix elements.

In the PS gauge, the currents are obtained by the projection:
\begin{align}
  &J^\mu_{\tt PSG}(k,k';\sig,\sig') \notag\\
  &=\gmntu\, J_\nu(k,k';\sig,\sig') \notag\\
  &=\sum_{\lam=\pm,0}(-\eta_\lam)\,\ept^{\mu}(q,\lam)\,\ept^{\nu}(q,\lam)^*
  J_\nu(k,k';\sig,\sig') \notag\\
  &=\sum_{\lam=\pm,0} {\cal J}_{\sig\sig'}^\lam(k,k') \,\ept^\mu(q,\lam). 
\label{J_psg}  
\end{align}
It is worth noting that the splitting amplitudes~\eqref{amp} are exactly the same between the PS gauge~\eqref{J_psg} and the Feynman gauge~\eqref{J_fg}
because of current conservation $q_\mu J^\mu=0$ for the currents of external on-shell quarks~\eqref{J_q} and gluons~\eqref{J_g}.
The PSG currents above are, however, free from large longitudinal component even in the highly boosted collinear region,
and hence the absolute value squared of the individual Feynman amplitude reproduces the parton-shower behavior in the collinear limit.
This shows that the severe cancellation among amplitudes in the original \helas code that adopts Feynman gauge does not exist in the PS gauge.

The splitting amplitudes for final-state radiations,
such as the time-like ($q^2>0$) $g^*\to q\bar q$ and $g^*\to gg$ currents,
have the same form as Eq.~\eqref{J_psg}, with $\eta_\lam=1$, in the PS gauge:
\begin{align}
  J^\mu_{\tt PSG}(k,k';\sig,\sig') 
 =\sum_{\lam=\pm,0} {\cal J}_{\sig\sig'}^\lam(k,k') \,\ept^\mu(q,\lam)^*
\end{align}
with
\begin{align}
   {\cal J}_{\sig\sig'}^\lam(k,k') = -J_\mu(k,k';\sig,\sig')\,\eps^\mu(q,\lam) 
\label{amp_f}   
\end{align}
when $q^2=(k+k')^2>0$.
The boost-invariant splitting amplitudes ${\cal J}_{\sig\sig'}^\lam(k,k')$ have the same form as obtained in ref.~\cite{Hagiwara:2009wt}, and the off-shell current is free from large components when the virtual gluon is highly boosted.

We note here that the splitting amplitudes~\eqref{amp} and \eqref{amp_f} for $g\to gg$ are not boost invariant for the currents calculated by \helas codes, 
{\tt JGGXXX}~\cite{Hagiwara:1990dw,Murayama:1992gi}.
This is because the massless gauge boson wave functions in \helas are fixed by the light-cone gauge with $n^\mu=(1,-\vec k/|\vec k|)$ by using the three momentum $\vec k$ of each external photons and gluons.
Although this particular choice of the external massless gauge boson wave functions allows \helas to obtain helicity amplitudes without specifying a particular gauge vector, the splitting amplitudes~\eqref{amp} and \eqref{amp_f} evaluated by \helas codes do not have the boost-invariant form of ref.~\cite{Hagiwara:2009wt}, whose absolute value squares are the splitting functions of the DGLAP equations.   
Nevertheless, the \helas currents have the splitting amplitude form of ref.~\cite{Hagiwara:2009wt} in the collinear limit, 
because the three momenta of external gluons are aligned with the gauge vector~\eqref{nvec} of the off-shell gluon.
This allows us to express not only the $q\to qg^*$ and $g^*\to q\bar q$ but also the $g\to gg^*$ and $g^*\to gg$ currents in terms of the boost-invariant splitting amplitudes of ref.~\cite{Hagiwara:2009wt} in the collinear limit. 

Summing up, in the PS gauge, the scalar component of the longitudinal state, which behaves badly in the collinear limit, is subtracted systematically, 
and hence the absolute value squared of each Feynman amplitude reduces to the corresponding parton shower in the collinear limit.
The amplitudes are exact even away from the collinear limit, and hence all the interference among different Feynman amplitudes are reproduced exactly, as interference among corresponding parton-shower amplitudes.

Before turning to numerical demonstrations in the PS gauge, 
we briefly mention the special \helas subroutine {\tt JEEXXX}, which was introduced to evaluate collinear photon emissions in $e^+e^-$ collisions~\cite{Hagiwara:1990dw,Murayama:1992gi}.
In order to accurately compute the current with nearly on-shell photon propagator, 
the original \helas current {\tt JIOXXX} is shifted by a term proportional to its four momentum $q^\mu$:
\begin{align}
  J_{\tt JEE}^\mu=J_{\tt JIO}^{\mu} - \frac{J_{\tt JIO}^{0}}{q^{0}}q^\mu.
\label{JEE}
\end{align}
When $|q^2|\ll 1$, the $\mu=0$ and 3 components of the current $J_{\tt JIO}^{\mu}$ grow as $q^0/\sqrt{|q^2|}$ 
because of the longitudinal component.
The subtraction~\eqref{JEE} suppresses these large components which cause numerical instability because of subtle cancellation among different Feynman amplitudes. 
In the PS gauge, the current is expressed as
\begin{align}
  J_{\tt PSG}^\mu &=\tilde g^{\mu\nu}{J_{\tt JIO}}_\nu 
  =J_{\tt JIO}^\mu-\frac{n\cdot J_{\tt JIO}}{n\cdot q}q^\mu 
  =J_{\tt JIO}^\mu-\frac{J_{\tt JIO}^0}{|\vec q|}q^\mu,
\end{align}
which differs from the {\tt JEE} current only by a term which vanishes in the small $|q^2|$ limit.

\section{Sample results}\label{sec:results}

In order to demonstrate the power of the PS gauge in numerical calculation of the amplitudes, 
we study two representative partonic subprocesses:
\begin{align}
  u(k_1)+d (k_2)&\to u(k_3)+d(k_4)+g(k_5), \label{p1}\\
  g(k_1)+g(k_2) &\to g(k_2)+g(k_4)+g(k_5), \label{p2}
\end{align}
in the Feynman gauge (with original {\tt HELAS}) and in the PS gauge with the new \helas codes presented in Appendix. 

For both processes,
event samples are generated with head-on collisions of initial partons at 2 \TeV\ centre-of-mass energy
with the minimal kinematical cuts for the final-state particles:
\begin{align}
 &p_{Ti}>20~{\rm GeV},\\ 
 &\Delta R_{ij}=\sqrt{(\eta_{i}-\eta_{j})^{2} + (\phi_{i} - \phi_{j})^{2}}>0.4.
\end{align}
We also introduce the hard-scattering scale as
\begin{align}
  \max(p_{Ti})>200\ {\rm GeV},
\end{align}
which allows us to interpret the amplitudes in the parton-shower language,
namely as a 2-to-2 hard process of the scale greater than 200~GeV times a parton splitting of the scale down to $\kTij$ of 8~GeV ($=20~{\rm GeV}\times0.4$).
Here, the relative transverse momentum between final-state particles, $\kTij$, is defined as
\begin{align}
  k_{T ij} = \min( \pT_i, \pT_j)\, \Delta R_{ij}.
\end{align}

\begin{figure}
\includegraphics[width=1.\columnwidth]{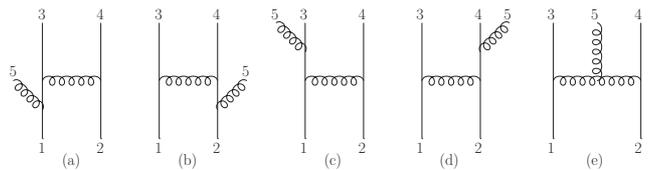} 
\caption{Feynman diagrams of the process, $u(1)d(2)\to u(3)d(4)g(5)$.}
\label{fig:diagram_ud_udg}
\end{figure}

Feynman diagrams of the process~\eqref{p1} are shown in Fig.~\ref{fig:diagram_ud_udg}.
They are categorized into these five types according to gluon radiation topologies; 
a gluon is radiated from one of the initial-state particles in (a) and (b),
from one of the final-state particles in (c) and (d), 
and from the internal propagator in (e). 

\begin{figure*}
\center
\includegraphics[width=0.75\textwidth]{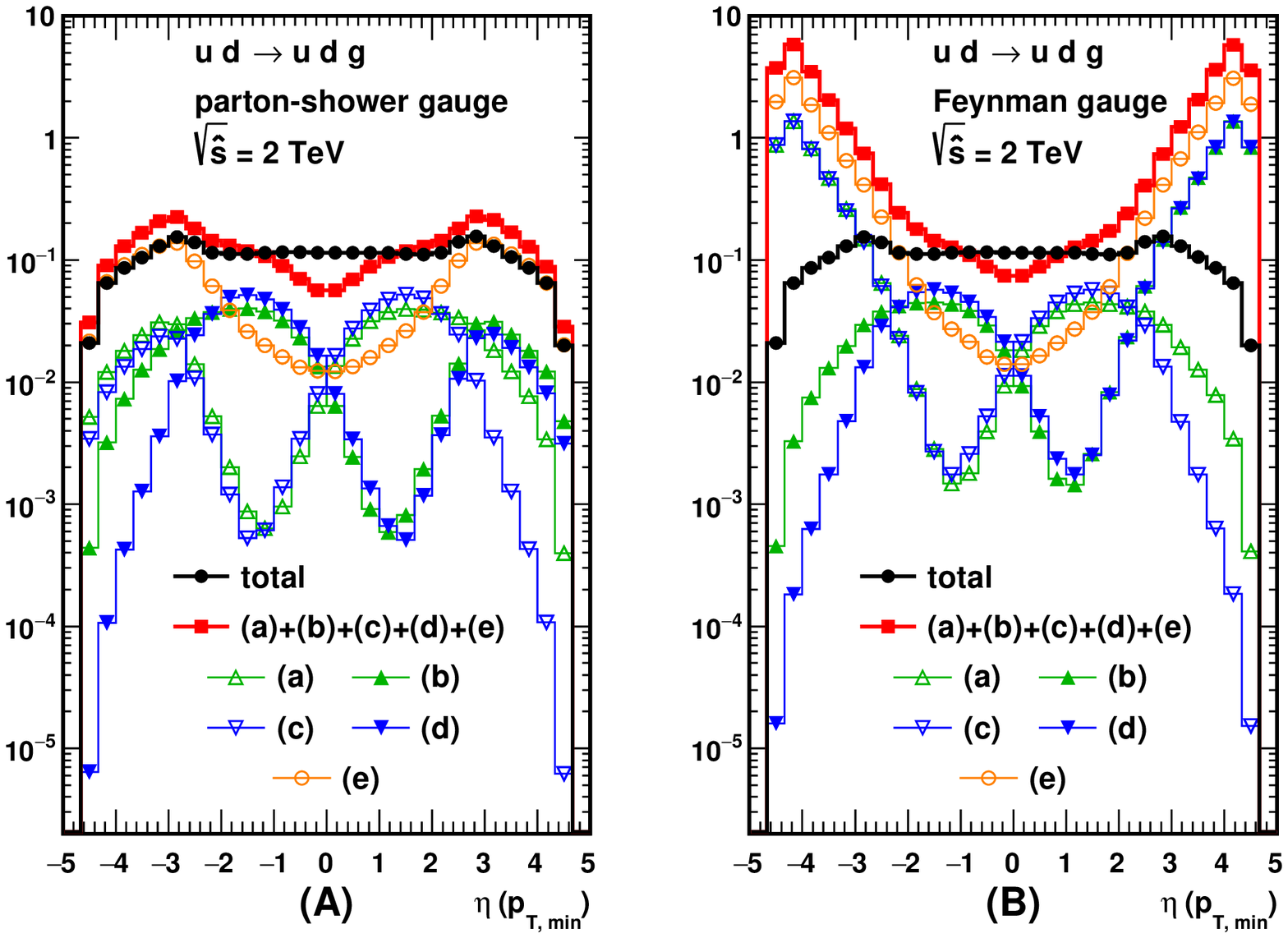}
\caption{Distributions of pseudorapidity of the final-state particle with minimum transverse momentum for the process $ud\to udg$ in the PS gauge (A) and in the Feynman gauge (B).  
A line with filled circles denotes the total distribution, 
while lines with triangles (a-d) and open circles (e) show the distribution of the squared amplitude of each type of the Feynman diagrams depicted in Fig.~\ref{fig:diagram_ud_udg}.  
A line with filled squares presents the distribution of the sum of the squared amplitudes of each diagram.
}
\label{fig:ud_udg_eta_pt_min}
\end{figure*}

In Fig.~\ref{fig:ud_udg_eta_pt_min}, we show the pseudorapidity of a final-state particle with minimum transverse momentum, \etapTmin, 
where the result in the PS gauge are plotted in the left-hand side, (A), and that in the Feynman gauge are plotted in the right-hand side, (B).
A line with filled circles denotes the differential cross section, i.e. the absolute value squared of the sum of all the amplitudes. 
On the other hand, lines with triangles (a-d) and open circles (e) show the distribution of the square of each Feynman amplitude depicted in Fig.~\ref{fig:diagram_ud_udg} (a) to (e), respectively.  
A line with filled squares presents the distribution of the sum of the squared amplitudes of each diagram, i.e. without any interference effects among the amplitudes.

\begin{figure*}
\center
\includegraphics[width=0.75\textwidth]{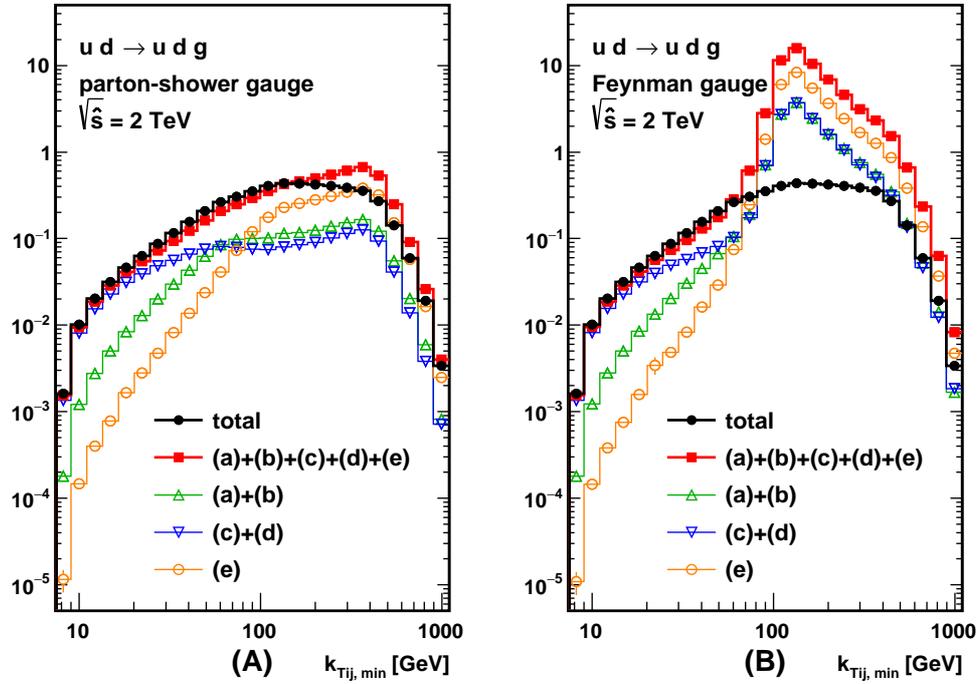}
\caption{Same as Fig.~\ref{fig:ud_udg_eta_pt_min}, but for distributions of minimum relative transverse momentum between final-state particles.}
\label{fig:ud_udg_logktij_min}
\end{figure*}

The total differential distributions, denoted by lines with filled circles, are identical between the PS gauge (A) and the Feynman gauge (B) as expected.
However, the contributions from each amplitude are quite different between the two gauges, especially in the large $\eta$ region. 
In the Feynman gauge, each squared amplitude in the large $\eta$ region becomes large due to the singularities, causing subtle cancellation of up to two orders of magnitudes among the amplitudes.
On the other hand, in the central region $|\eta|<2$, where the contribution from the soft gluon is dominant, 
all the diagrams contribute almost equally and the contributions from each diagram in the two gauges
are very similar. 
We note that the flat $\eta$ distribution is a consequence of boost invariance of the soft-gluon emission amplitudes.  

In order to study the splitting in the initial state, let us focus on the large $\eta$ region in Fig.~\ref{fig:ud_udg_eta_pt_min}(A),
where contributions can be explained by the hard-scattering process and the splitting probability in the corresponding parton-shower history.   
The large positive $\eta$ jet with the smallest $p_T$ is mostly $u(3)$ from $u(1)\to u(3)+g^*$ splitting which are common in diagrams (b), (d) and (e). 
The hard process is then $g^* + d(2) \to d(4) + g(5)$.   
In the on-shell limit of the virtual gluon $g^*$, the sum of the three hard-scattering amplitudes give the physical amplitudes for $gd\to gd$.
The destructive interference observed in the large $\eta$ region of Fig.~\ref{fig:ud_udg_eta_pt_min}(A) hence shows the interference among the three amplitudes of the hard-scattering process.   
In the PS gauge, therefore, one can even observe the interference pattern among hard-scattering amplitudes. 
Similarly, the large negative $\eta$ jet with the smallest $p_T$ is mostly $d(4)$ from 
$d(2)\to d(4)+g^*$ splitting which are common in diagrams (a), (c) and (e), 
and the same argument can be applied.

Next, in order to study the final-state splitting, we show the minimum relative transverse momentum between final-state particles, \kTijmin, in Fig.~\ref{fig:ud_udg_logktij_min}.
As in Fig.~\ref{fig:ud_udg_eta_pt_min}, the results in PS gauge are shown in the left panel (A), while those in Feynman gauge are shown in the right (B).
Lines with filled circles show the differential cross section obtained by squaring the sum of all the amplitudes, and hence they are identical between (A) and (B).
The sum of the square of each diagram is shown by a line with filled squares, which shows subtle cancellation among amplitudes at large \kTijmin\ ($\gtrsim 100~\GeV$) in the Feynman gauge (B).

On the other hand, at small \kTijmin\ ($\lesssim 30~\GeV$), we can observe the parton-shower behavior, where the hard-scattering (with $\pT > 200~\GeV$) process $ud \to ud$ is followed by collinear gluon emission from final-state $u$ or $d$, corresponding to the Feynman diagrams (c) and (d), respectively, in Fig.~\ref{fig:diagram_ud_udg}.
Because the contribution of the square of the amplitude (c) and that of (d) are identical, we show their sum by a line with inverted triangles.
Likewise the sum of squares of the diagrams (a) and (b) are shown by a line with normal triangles.
The dominance of the contribution of the diagrams (c) and (d) at small \kTijmin\ is observed even in the Feynman gauge (B), 
because the virtual $u$ and $d$ propagators in (c) and (d) are common between the two gauges. 

\begin{figure}
\includegraphics[width=1.\columnwidth]{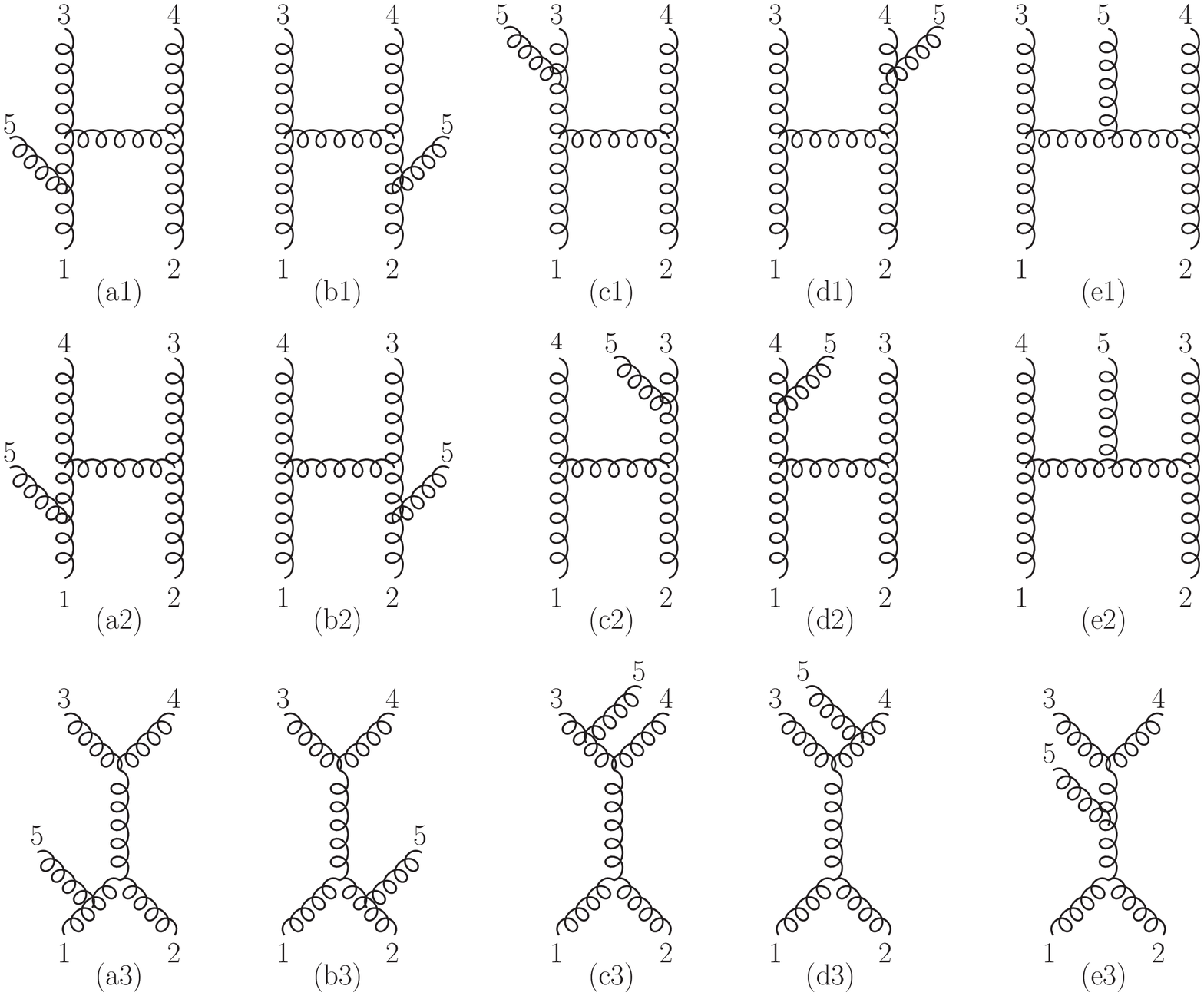}
\caption{Feynman diagrams of the process, $g(1)g(2)\to g(3)g(4)g(5)$.}
\label{fig:diagram_gg_ggg}
\end{figure}

Now let us turn to the process~\eqref{p2}, whose Feynman diagrams are depicted in Fig.~\ref{fig:diagram_gg_ggg}.
Similar to $ud\to udg$, the diagrams are categorized into five types.
In the type (a), the gluon (5) is emitted from $g(1)$.
Likewise $g(5)$ is emitted from $g(2)$ in (b), a virtual gluon splits into $g(3)$ and $g(5)$ in (c), or into $g(4)$ and $g(5)$ in (d).
In the diagrams of type (e), $g(5)$ is emitted from an exchanged virtual gluon.
For each type, we have $t$-channel (a1)-(e1), $u$-channel (a2)-(e2), as well as the $s$-channel gluon exchange diagrams (a3)-(e3).
In addition, although we do not show explicitly in Fig.~\ref{fig:diagram_gg_ggg}, there are diagrams with the four-point vertex.

\begin{figure*}
\center
\includegraphics[width=0.75\textwidth]{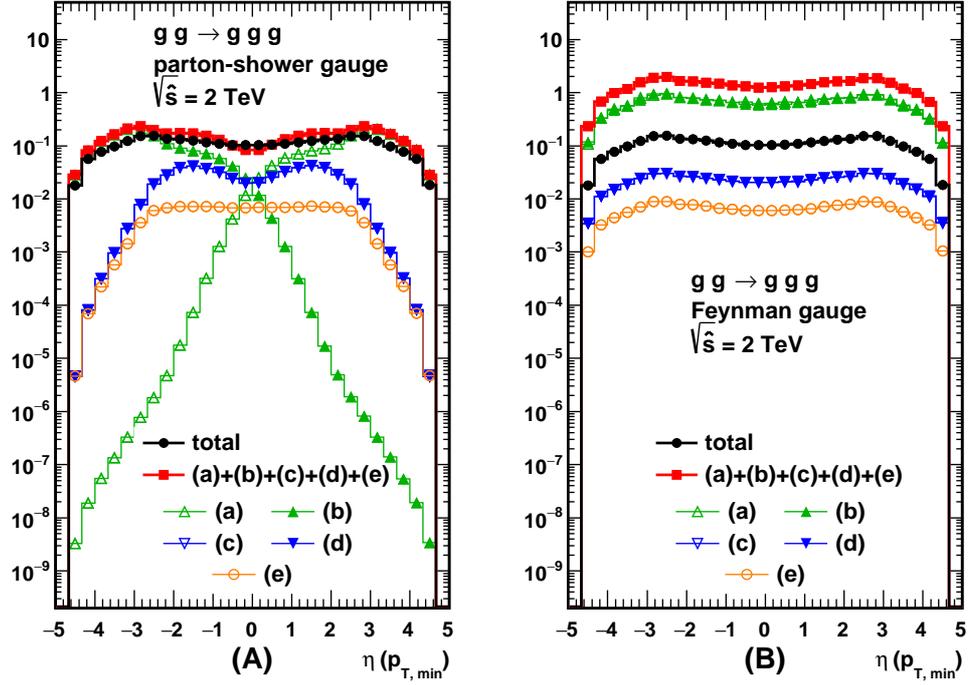}
\caption{
Distributions of pseudorapidity of the final-state particle with minimum transverse momentum for the process $gg\to ggg$ in the PS gauge (A) and in the Feynman gauge (B).  
A line with filled circles denotes the total distribution, 
while lines with triangles (a-d) and open circles (e) show the distribution of the squared amplitudes of each type of the Feynman diagrams depicted in Fig.~\ref{fig:diagram_gg_ggg}.  
A line with filled squares presents the distribution of the sum of the squared amplitudes of each diagram.
}
\label{fig:gg_ggg_eta_pt_min}
\end{figure*}

\begin{figure*}
\center
\includegraphics[width=0.75\textwidth]{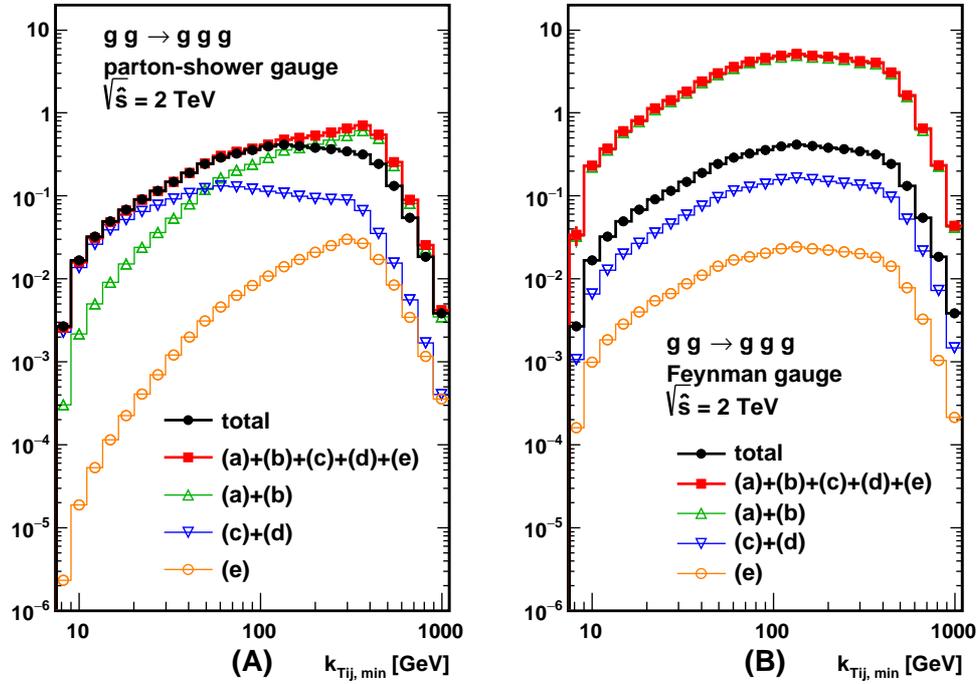}
\caption{Same as Fig.~\ref{fig:gg_ggg_eta_pt_min}, but for distributions of minimum relative transverse momentum between final-state particles.
}
\label{fig:gg_ggg_logktij_min}
\end{figure*}

Similar to Figs.~\ref{fig:ud_udg_eta_pt_min} and \ref{fig:ud_udg_logktij_min}, we present results for $gg\to ggg$ in Figs.~\ref{fig:gg_ggg_eta_pt_min} and \ref{fig:gg_ggg_logktij_min}. 
The differential cross sections shown by lines with filled circles are given by the absolute value squared of the sum of all the amplitudes including the diagrams with the four-point vertex, and they agree in both gauges as expected.
The curves with triangles (open and filled, normal and inverted) and open circles are obtained by 
squaring the sum of the three diagrams in each column of Fig.~\ref{fig:diagram_gg_ggg},
such as (a1)+(a2)+(a3) for the type (a), 
(b1)+(b2)+(b3) for the type (b), etc.
We ignore amplitudes with the four-point vertex in these subset of diagrams, since their magnitudes are always found to be small.

In Fig.~\ref{fig:gg_ggg_eta_pt_min}(A), the large positive $\eta$ jet comes from the diagrams of type (a), shown by open normal triangles, while the large negative $\eta$ jet comes from the diagrams of type (b), shown by filled normal triangles.
In case of (a), the smallest $p_T$ jet is $g(5)$, emitted from $g(1)$, 
and therefore the hard process is $g^*+g(2)\to g(3) + g(4)$.  
Similarly, for the large negative $\eta$ region, 
the smallest $p_T$ jet is $g(5)$, emitted from $g(2)$, 
and the hard process is $g(1)+g^*\to g(3) + g(4)$.   
Even in the central region $|\eta|<2$, where soft gluons with wide angles mostly contribute, the above hard processes are dominant, except  
at the very central region $|\eta|\sim0$, where all the diagrams contribute almost equally.
In the PS gauge the sum of the absolute value squared of each amplitude can reproduce the physical distribution with small interference effects.

In contrast, in Fig.~\ref{fig:gg_ggg_eta_pt_min}(B), 
all the sub-diagrams have the same behavior; 
(a) and (b) give the same symmetric $\eta$ distribution, so do (c) and (d).
This implies that, in the Feynman gauge, the direction of the virtual gluon three momentum cannot be distinguished
and the singular behavior is shared by all the diagrams.
The total sum of the squares of the five subamplitudes is given by filled squares, which agrees approximately with the differential cross section in the PS gauge (A), whereas it is one order of magnitude larger than the cross section in the Feynman gauge (B).

In Fig.~\ref{fig:gg_ggg_logktij_min}(A), similar to the $ud\to udg$ case in Fig.~\ref{fig:ud_udg_logktij_min}(A), 
the dominance of the final-state splitting (c)+(d) can clearly be seen in the low \kTijmin\ ($\lesssim 30~\GeV$) region.
This reproduces the corresponding parton-shower histories, where the hard scattering gives $g(1)+g(2) \to g(3)+g(4)$, followed by final-state collinear $g(5)$ emission from $g(3)$ (c) or $g(4)$ (d).
In contrast, in the Feynman gauge in Fig.~\ref{fig:gg_ggg_logktij_min}(B), the contribution from the initial-state splitting (a)+(b) is much larger than that from the final-state splitting (c)+(d), even in the small \kTijmin\ region.
Unlike the final-state splitting $q^* \to qg$, whose collinear-splitting behavior can be observed in both gauges in Fig.~\ref{fig:ud_udg_logktij_min}, the virtual gluon splitting $g^* \to gg$ depends strongly on the gauge because of the large longitudinal component in the collinear limit.


\section{Summary}\label{sec:summary}

We proposed a new propagator form of photons and gluons,
in which the scalar component of the longitudinal polarization contribution is subtracted systematically by making use of the BRST invariance of the off-shell gauge boson created from physical on-shell states.
The propagator has the light-cone gauge form with the gauge vector pointing along the opposite of the off-shell photon or gluon three momentum.
We may call the gauge as parton-shower (PS) gauge since 
in collinear configurations
the absolute value squared of each Feynman amplitude reproduces all the singular behaviors of the corresponding parton shower in this gauge.
We showed in simple $2\to 3$ subprocesses that 
calculations in the PS gauge can avoid subtle gauge cancellations among Feynman amplitudes that appear in the Feynman gauge.
Because of these properties, we believe that amplitudes evaluated in the PS gauge give us deeper insights on scattering processes, and possibly give improved efficiency in event generations,
especially with single-diagram-enhanced multi-channel integration~\cite{Maltoni:2002qb}.

The new \helas subroutines in the PS gauge are listed in Appendix, and they are publicly available on the web site.%
\footnote{HELAS/MadGraph/MadEvent Home Page:\\
{\tt https://madgraph.ipmu.jp/IPMU/}.}
Numerical calculations in the PS gauge can also be done with {\tt MadGraph5\_aMC@NLO} version 2.7.0 or later.

\section*{Acknowledgements}
We would like to thank Olivier Mattelaer for the implementation of the PS gauge into {\tt MadGraph5\_aMC@NLO}. 
The work of K.M. was supported in part by JSPS KAKENHI Grant No.~18K03648.

\appendix
\def\thesection{Appendix \Alph{section}}

\section{Sample codes in the parton-shower gauge}\label{sec:codes}

In this section we list \helas subroutines that are needed to compute off-shell vector currents in the PS gauge.
For future extension of \helas we modify the internal structure of \helas wave functions.
In the original \helas wave functions the four momentum of particles is stored in the last two components of the complex six-dimensional array.
In new \helas which includes new PS-gauge calculation the first two components of the wave functions contain the four momentum%
\footnote{{\tt ALOHA}~\cite{deAquino:2011ub} employs this structure.}
because the extension of the array of wave functions becomes possible to include further information.
We note that the original \helas library was coded by Fortran77, while the new library is coded by Fortran90.

\subsection{Parameter file and utility subroutine}
New parameters to control helicity amplitude calculation are included in the new \helas library.
In \texttt{helas\_param.f90} (List~\ref{list:helas_param}) a new flag, \texttt{helas\_mode}, is introduced.
Feynman gauge is used when ${\tt helas\_mode}=1$ and PS gauge is used when ${\tt helas\_mode}=2$ for massless gauge boson propagators.
This flag can be changed by calling \texttt{hlmode({\it helas\_mode})} (List~\ref{list:hlmode}), then all the gauge definition for massless gauge-boson propagators is changed.

\begin{lstlisting}[caption=helas\_param.f90,label=list:helas_param]
module helas_param

  implicit none

  integer, save :: helas_mode = 1
  integer, parameter :: helas_mode_feynman = 1
  integer, parameter :: helas_mode_psg = 2
  integer, parameter :: n_helas_mode = 2

end module helas_param
\end{lstlisting}

\begin{lstlisting}[caption=hlmode.f90,label=list:hlmode]
subroutine hlmode(mode)

  use helas_param

  implicit none

  integer, intent(in) :: mode

  integer :: i
  

  if (mode>0 .and. mode<=n_helas_mode) then

     helas_mode = mode

  endif
  
end subroutine hlmode
\end{lstlisting}

\subsection{Off-shell vector-current subroutines}
We introduce subroutines to compute off-shell massless vector-boson currents in the PS gauge that are used in this paper.
The list of the arguments in these subroutines is the same as that in the original \helas library~\cite{Murayama:1992gi}, so that it is not necessary to modify amplitude computation program to use PS gauge instead of Feynman gauge.

\subsubsection{jioxxx}
The function, \texttt{jioxxx} (List~\ref{list:jioxxx}), computes an off-shell vector current attached to an external fermion line.
The vector-boson propagator is given in the PS gauge for a massless vector if the flag, \texttt{helas\_mode}, is set to 2.

\begin{lstlisting}[caption=jioxxx.f90,label=list:jioxxx]
subroutine jioxxx(fi,fo,gc,vmass,vwidth , jio)

  use helas_param
  
  implicit none

  double complex, intent(in) :: fi(6),fo(6)
  double complex, intent(in) :: gc(2)
  double precision, intent(in) :: vmass, vwidth
  double complex, intent(out) :: jio(6)

  double complex :: jj(0:4)
  double complex :: j12(0:4)
  double complex :: js,d
  double complex :: js1,js2
  double complex :: q(0:4)
  double precision :: qabs2,qabs
  double precision :: q2,vm2
  double complex :: cm2 

  double precision :: n(0:4)
  
  double precision :: nq

  double complex, parameter :: ci = (0.d0, 1.d0)
  

  jio(1:2) = fo(1:2)-fi(1:2)

  q(0) = -dble( jio(1))
  q(1) = -dble( jio(2))
  q(2) = -dimag(jio(2))
  q(3) = -dimag(jio(1))
  q(4) = -ci*vmass
  qabs2 = dble(q(1))**2+dble(q(2))**2+dble(q(3))**2
  qabs = sqrt(qabs2)
  q2 = dble(q(0))**2-qabs2

  vm2 = vmass**2

  j12(0) =  gc(1)*( fo(5)*fi(3)+fo(6)*fi(4)) &
           +gc(2)*( fo(3)*fi(5)+fo(4)*fi(6))
  j12(1) = -gc(1)*( fo(5)*fi(4)+fo(6)*fi(3)) &
           +gc(2)*( fo(3)*fi(6)+fo(4)*fi(5))
  j12(2) =( gc(1)*( fo(5)*fi(4)-fo(6)*fi(3)) &
           +gc(2)*(-fo(3)*fi(6)+fo(4)*fi(5)))*ci
  j12(3) =  gc(1)*(-fo(5)*fi(3)+fo(6)*fi(4)) &
           +gc(2)*( fo(3)*fi(5)-fo(4)*fi(6))

  d = 1.d0/dcmplx( q2-vm2, vmass*vwidth )

  if (vmass/=0.d0) then

     cm2=dcmplx( vm2, -vmass*vwidth )

     js = (q(0)*j12(0)-q(1)*j12(1)- &
          q(2)*j12(2)-q(3)*j12(3)) / cm2

     jj(0:3) = j12(0:3) - q(0:3)*js

  else

     jj(0:3) = j12(0:3)
     
  end if

  jio(3:6) = jj(0:3) * d

  if (helas_mode==helas_mode_psg) then

     if (vmass==0.d0) then

        if (qabs2>=dble(q(0))**2) then
           n(0) = sign(1.d0,dble(q(0)))
           n(1) = -dble(q(1))/qabs
           n(2) = -dble(q(2))/qabs
           n(3) = -dble(q(3))/qabs
        else
           n(0) = 1.d0
           n(1) = 0.d0
           n(2) = 0.d0
           n(3) = -1.d0
        endif

        nq = n(0)*dble(q(0))-n(1)*dble(q(1)) &
             -n(2)*dble(q(2))-n(3)*dble(q(3))

        js1 = (n(0)*j12(0)-n(1)*j12(1) &
             -n(2)*j12(2)-n(3)*j12(3)) / nq

        js2 = (q(0)*j12(0)-q(1)*j12(1) &
             -q(2)*j12(2)-q(3)*j12(3)) / nq
        
        jio(3:6) = (j12(0:3)-q(0:3)*js1-n(0:3)*js2) * d
        
     endif
     
  end if

end subroutine jioxxx
\end{lstlisting}

\subsubsection{jggxxx}
The function, \texttt{jggxxx} (List~\ref{list:jggxxx}), computes an off-shell gluon current from the three-point gluon vertex attached with the gluon propagator depending on the \texttt{helas\_mode} flag.

\begin{lstlisting}[caption=jggxxx.f90,label=list:jggxxx]
subroutine jggxxx(v1,v2,g, jgg)

  use helas_param
  
  implicit none

  double complex, intent(in) :: v1(6),v2(6)
  double precision, intent(in) :: g
  double complex, intent(out) :: jgg(6)

  double complex :: j12(0:3),v12
  double precision :: p1(0:3),p2(0:3),q(0:3),q2
  double precision :: d

  double complex :: p23v1, p31v2
  double complex :: js1, js2
  
  double precision :: qabs,qabs2
  double precision :: nq
  double precision :: n(0:3)
  

  jgg(1:2) = v1(1:2) + v2(1:2)

  p1(0) = dble( v1(1))
  p1(1) = dble( v1(2))
  p1(2) = dimag(v1(2))
  p1(3) = dimag(v1(1))

  p2(0) = dble( v2(1))
  p2(1) = dble( v2(2))
  p2(2) = dimag(v2(2))
  p2(3) = dimag(v2(1))
  
  q(0) = -(p1(0)+p2(0))
  q(1) = -(p1(1)+p2(1))
  q(2) = -(p1(2)+p2(2))
  q(3) = -(p1(3)+p2(3))

  qabs2 = q(1)**2+q(2)**2+q(3)**2
  qabs = sqrt(qabs2)

  q2 = dble(q(0))**2-qabs2

  v12 = v1(3)*v2(3)-v1(4)*v2(4)-v1(5)*v2(5)-v1(6)*v2(6)
  
  p23v1 = (p2(0)-q(0))*v1(3) - (p2(1)-q(1))*v1(4) &
        - (p2(2)-q(2))*v1(5) - (p2(3)-q(3))*v1(6)
  p31v2 = (q(0)-p1(0))*v2(3) - (q(1)-p1(1))*v2(4) &
        - (q(2)-p1(2))*v2(5) - (q(3)-p1(3))*v2(6)

  j12(0:3) = -g*((p1(0:3)-p2(0:3))*v12 + p23v1*v2(3:6) + &
       p31v2*v1(3:6))

  d = 1.d0/q2
  
  jgg(3:6) = j12(0:3) * d

  if (helas_mode==helas_mode_psg) then

     if (qabs2>=dble(q(0))**2) then
        n(0) = sign(1.d0,dble(q(0)))
        n(1) = -dble(q(1))/qabs
        n(2) = -dble(q(2))/qabs
        n(3) = -dble(q(3))/qabs
     else
        n(0) = 1.d0
        n(1) = 0.d0
        n(2) = 0.d0
        n(3) = -1.d0
     endif

     nq = n(0)*dble(q(0))-n(1)*dble(q(1))- &
          n(2)*dble(q(2))-n(3)*dble(q(3))

     js1 = (n(0)*j12(0)-n(1)*j12(1) &
          -n(2)*j12(2)-n(3)*j12(3)) / nq
     
     js2 = (q(0)*j12(0)-q(1)*j12(1) &
          -q(2)*j12(2)-q(3)*j12(3)) / nq
     
     jgg(3:6) = (j12(0:3)-q(0:3)*js1-n(0:3)*js2) * d
        
  endif
  
end subroutine jggxxx
\end{lstlisting}

\subsubsection{jgggxx}
The function, \texttt{jgggxx} (List~\ref{list:jgggxx}), computes an off-shell gluon current from the fourt-point gluon vertex attached with the gluon propagator depending on the \texttt{helas\_mode} flag.

\begin{lstlisting}[caption=jgggxx.f90,label=list:jgggxx]
subroutine jgggxx(w1,w2,w3,g, jggg)

  use helas_param
  
  implicit none

  double complex, intent(in) :: w1(6),w2(6),w3(6)
  double precision, intent(in) :: g
  double complex, intent(out) :: jggg(6)
  
  double complex :: dv1(0:3),dv2(0:3),dv3(0:3)
  double complex :: j123(0:3),d,v23,v31
  double complex :: js1,js2
  double precision :: p1(0:3),p2(0:3),p3(0:3),q(0:3),g2,q2

  double precision, parameter :: rZero = 0.d0
  double precision, parameter :: rOne = 1.d0

  double precision :: qabs, qabs2
  double precision :: nq
  double precision :: n(0:3)
      
 
  jggg(1:2) = w1(1:2)+w2(1:2)+w3(1:2)

  p1(0) = dble(      w1(1))
  p1(1) = dble(      w1(2))
  p1(2) = dimag(w1(2))
  p1(3) = dimag(w1(1))

  p2(0) = dble(      w2(1))
  p2(1) = dble(      w2(2))
  p2(2) = dimag(w2(2))
  p2(3) = dimag(w2(1))

  p3(0) = dble(      w3(1))
  p3(1) = dble(      w3(2))
  p3(2) = dimag(w3(2))
  p3(3) = dimag(w3(1))

  q(0) = -(p1(0)+p2(0)+p3(0))
  q(1) = -(p1(1)+p2(1)+p3(1))
  q(2) = -(p1(2)+p2(2)+p3(2))
  q(3) = -(p1(3)+p2(3)+p3(3))
  
  qabs2 = dble(q(1))**2+dble(q(2))**2+dble(q(3))**2
  qabs = sqrt(qabs2)

  q2 = dble(q(0))**2-qabs2
      
  dv1(0:3) = w1(3:6)
  dv2(0:3) = w2(3:6)
  dv3(0:3) = w3(3:6)

  g2 = g**2

  d = 1.d0/dcmplx( q2 )

  v23 = dv3(0)*dv2(0)-dv3(1)*dv2(1)- &
       dv3(2)*dv2(2)-dv3(3)*dv2(3)
  v31 = dv1(0)*dv3(0)-dv1(1)*dv3(1)- &
       dv1(2)*dv3(2)-dv1(3)*dv3(3)

  j123(0:3) = g2*( v23*dv1(0:3) - v31*dv2(0:3) )

  jggg(3:6) = j123(0:3) * d

  if (helas_mode==helas_mode_psg) then
     
     if (qabs2>=dble(q(0))**2) then
        n(0) = sign(1.d0,dble(q(0)))
        n(1) = -dble(q(1))/qabs
        n(2) = -dble(q(2))/qabs
        n(3) = -dble(q(3))/qabs
     else
        n(0) = 1.d0
        n(1) = 0.d0
        n(2) = 0.d0
        n(3) = -1.d0
     endif

     nq = n(0)*dble(q(0))-n(1)*dble(q(1))- &
          n(2)*dble(q(2))-n(3)*dble(q(3))

     js1 = (n(0)*j123(0)-n(1)*j123(1)- &
          n(2)*j123(2)-n(3)*j123(3)) / nq

     js2 = (q(0)*j123(0)-q(1)*j123(1)- &
          q(2)*j123(2)-q(3)*j123(3)) / nq
     
     jggg(3:6) = (j123(0:3)-q(0:3)*js1-n(0:3)*js2) * d

  endif
  
end subroutine jgggxx
\end{lstlisting}

\vspace*{1cm}
\bibliographystyle{JHEP}
\bibliography{bibpsg}


\end{document}